**Exchange for Growth: Currency Dynamics in Emerging Markets**

*Rohan Dubey* and *Shaunak Kulkarni*

October 08, 2025

Working paper for public distribution.

Research originally produced for MDAE YERC (9[th] edition) on January 31, 2025.

Exchange for Growth: Currency Dynamics in Emerging Markets

**Introduction**

The modern archetype of a fast-growing, super-heated Emerging Market Economy (EME) operating at full capacity gained popularity in the decades following the East Asian miracle, and has come to represent an ideal path to economic maturity that often overlooks the fragility of an export-oriented industrial base. The exemplary success of East and Southeast Asian economies in the late 20$^{th}$ century is as much a product of effective policy and informed economic activity, as it is the result of favourable economic trends and changing technology (Stiglitz, 1996); it is an exemplar of the best-case outcome of highly leveraged growth, the hazards of which were demonstrated through financial crises that spread across emerging markets in Asia and Latin America (Sachs, et al., 1996). Many countries have tried to replicate the hot-growth approach of the East Asian miracle with varying degrees of success, and a recurring theme across EMEs which we find worth investigating (in the context of currency dynamics) has been the peculiar reaction to domestic monetary policy that is theorised to be a consequence of phenomena known as the "price puzzle" and the "FX puzzle" acting in parallel.

An empirically substantiated hypothesis explaining the unconventional outcome involves accounting for expectations in terms of inflation, growth, and exchange rate, which alludes to economically significant activity in transmission channels between policy authorities that affects growth and real capital flows (Ha, et al., 2024); from this, one may infer an implicit relationship between the character of EME growth and international capital flows, through which we seek to substantiate the understanding of globalised growth as an economic characteristic that is an endogenous factor to the equilibrium within an economy and between trade partners. In this study, we examine two plausible paths this balance follows to equilibrium can follow, focusing on propositions from the Mundell-Flemming open economy model as the basis for balance within an economy, and deriving the mechanics of inter-economy effects by means of monetary models for currency dynamics and capital flows.

## 1A – Assessing the Fiscal Vector: Key Factors & Methodology

The hypothesised impact of fiscal policy on the current account varies by schools of thought. Despite these differences, the Twin Deficits Hypothesis (TDH) remains a widely cited (and generally accepted) theoretical explanation of this relationship. While the Mundell-Fleming framework derives TDH from an examination of sectoral accounting flows, it critically assumes that the private savings gap remains either zero or stable. Given this necessary condition and conflicting empirical evidence (Banday & Aneja, 2019; K. G. & Gautam, 2015; Aloryito, et al., 2016; Koukouritakis & Panousis, 2020; Nautiyal, et al., 2023), we find it suitable to conduct an econometric assessment of the causal relationship between budget and current account deficits in the context of large monetarily sovereign developing economies, as an empirical starting point for a broader evaluation of the role of fiscal measures in shocks to currency valuation in emerging markets.

We undertake analysis for a representative set of developing nations (the selection of which is elaborated in Appendix I) with different underlying structures. Given the difficulty in establishing a causal relationship such as the one proposed by the Mundell-Flemming TDH, we break down (as detailed in Appendix II) the fundamental flow-through mechanism into a series of testable statistical hypotheses. We begin our analysis by outlining the series of hypotheses that will test the validity of TDH. Specifically, we will study:

1. The link between budget deficits, macroeconomic indicators, and interest rates in the Indian economy
2. Stability in the relationship between the long- and short-term rates in India
3. The predictive link between the change in budget deficits and current account deficit in the representative selection

Given the non-stationary and highly persistent nature of the control variables required, we use Autoregressive Distributed Lag (ARDL) models, as specified in Figure 1.

Exchange for Growth: Currency Dynamics in Emerging Markets| Independent Variable | Description | Rationale |
|---|---|---|
| *T-Bill 3-month (TB3M)* | Quarterly, Rate | Key Regressor (~authority) |
| *TB3M lag* | Quarterly (lagged), Rate | Regressor Control |
| *CPI* | Annual Change, Percent | Control (inflation/interest) |
| *Manufacturing Index (mind)* | Annual Change, Percent | Proxy (growth) |
| *Stock Market Index* | NIFTY Index (actual) | Proxy (growth) |
| *AR[1]* | First Lag | testing for persistence |
| *AR[2]* | Second Lag | testing for persistence |

~authority: in direct test for Monetary Authority/Sovereignty

*Figure 1: ARDL Model Composition*

Hypothesis 3 requires a more direct Granger causality test (Granger, 1969). Although we can directly proceed with Granger tests, the ARDL model explains the flow-through mechanism more comprehensively, and better supports an evaluation of economic implications.

## 1B – Assessing the Fiscal Vector: Outcomes & Inference

| 5-year bonds | coeff. | std. error | t-stat. | p-value |
|---|---|---|---|---|
| *constant* | -0.4055 | 0.873 | -0.464 | 0.643 |
| *CPI* | -0.0008 | 0.003 | -0.298 | 0.766 |
| *mind* | 0.0008 | 0.001 | 0.802 | 0.423 |
| *NIFTY* | -4.647e-06 | 9.23e-06 | -0.503 | 0.615 |
| *TB3M* | 0.4512 | 0.031 | 14.348 | 0.000 |
| *TB3M lag* | -0.3836 | 0.035 | -10.886 | 0.000 |
| *AR[1]* | 0.8725 | 0.052 | 16.741 | 0.000 |
| *AR[2]* | 0.0275 | 0.047 | 0.583 | 0.560 |
| *log_INRUSD* | 0.1969 | 0.232 | 0.849 | 0.397 |
| Obs. (Df)   261 (252) | | $R^2$   0.963 | | Adj. $R^2$   0.961 |
| F-stat.   810.9 | | AIC   -74.08 | | BIC   -42.00 |

*Table 1: ARDL statistics for 5-year yield in India*

Rohan Dubey & Shaunak Kulkarni                                                                                     Page 3 of 18

The statistical significance of the three-month treasury bills shows that the long-term yields on both the five- (Table 1) and ten-year (Table 2) bonds are a policy variable for the RBI. Given that these yields are a benchmark for other interest rates in the economy, statistical results show that the RBI has decisive influence on long-term interest rates.

| 10-year bonds | coeff. | std. error | t-stat. | p-value |
|---|---|---|---|---|
| constant | -0.4707 | 1.061 | -0.444 | 0.658 |
| CPI | -0.0012 | 0.003 | -0.391 | 0.696 |
| mind | 0.0027 | 0.001 | 2.308 | 0.022 |
| NIFTY | -3.767e-06 | 1.05e-05 | -0.357 | 0.721 |
| TB3M | 0.1358 | 0.039 | 3.502 | 0.001 |
| TB3M lag | -0.0501 | 0.041 | -1.228 | 0.220 |
| AR[1] | 0.7907 | 0.063 | 12.536 | 0.000 |
| AR[2] | 0.0541 | 0.061 | 0.890 | 0.374 |
| log_INRUSD | 0.2931 | 0.282 | 1.039 | 0.300 |
| Obs. (Df)  261 (252) | | $R^2$  0.925 | | Adj. $R^2$  0.922 |
| F-stat.  387 | | AIC  30.04 | | BIC  62.12 |

*Table 2: ARDL statistics for 10-year yield in India*

There is no evidence that other macroeconomic indicators such as exchange rates or the growth proxies used sway the long-term interest rates. Significantly, the table below shows the results from a Granger causality test that sought to identify the presence of any causal link between movements in the Indian government's fiscal deficit and the yield on ten-year government bonds.

| India (fiscal deficit on yield) | | |
|---|---|---|
| Lag | F-Statistic | p-Value |
| 1 | 0.0505 | **0.8243** |
| 2 | 0.3307 | **0.7225** |
| 3 | 0.4922 | **0.6927** |
| 4 | 0.2680 | **0.8933** |
| 5 | 0.2036 | **0.9536** |
| 6 | 0.1618 | **0.9794** |
| 7 | 0.1200 | **0.9915** |

*Table 3: Granger causality (fiscal deficit on yield) statistics for India*

The budget deficit does not lead to financial crowding out, contrary to what TDH would suggest. These results are consistent with empirical studies (Akram & Das, 2019) and theoretical arguments from economists across various schools of thought. Keynes was the first to posit that monetary authorities exert a decisive influence on long-term interest rates (Keynes, 1935); more recently, economists from both New Keynesian (Blanchard, 2019) and Post-Keynesian (Fullwiler, 2020; Lavoie, 2014) traditions have come to similar conclusions.

The second hypothesis is also confirmed by the statistical significance of TB3M lag and autoregressive terms. Clearly, there is a strong persistence in interest rates meaning that there is a stable link between short and long-term interest rates which policymakers can utilise to ensure that bond yields (and subsequently, general interest rates) remain consistent with objectives, regardless of movements in the foreign sector, macroeconomic conditions, and the government deficit.

A break-down of the financial crowding-out relationship implies the absence of a causal link between the current account deficit (CAD) and fiscal deficits. For a more comprehensive exposition of the complete flow-through mechanism of TDH, we provide global results from the Granger causality test in Appendix III, and the India case below.

| India (fiscal deficit on CAD) | | |
|---|---|---|
| *Lag* | *F-Statistic* | *p-Value* |
| 1 | 0.2892 | **0.5961** |
| 2 | 0.1994 | **0.8209** |
| 3 | 0.4793 | **0.7012** |
| 4 | 0.3122 | **0.8647** |
| 5 | 0.3609 | **0.8639** |
| 6 | 0.1783 | **0.9741** |
| 7 | 0.1072 | **0.9938** |

*Table 4: Granger causality (fiscal deficit on CAD) statistics for India*

Another reason for the rejecting TDH is its assumption that the private savings gap is stable. Empirical evidence (Argy, 1992) does not support this claim. Further, the national income identity:

$$(S - I) = (G - T) + NX$$

postulates that changes in the budget deficit can create changes in the private savings gap, thereby disrupting the causal link that TDH suggests. Factors such as capital controls, protectionist policy measures and other trade barriers can also disrupt the flow-through mechanism. Evidence shows that reliance on a TDH argument can lead to serious policy errors that exacerbate currency crises TDH may deter policymakers from investing towards import substitution of critical products such as food and electricity; doing so keeps developing nations dependent on volatile foreign exchange markets to source critical necessities, thereby increasing the pressure on the current account. In this sense, fiscal austerity supported by TDH may make a currency crisis self-reinforcing, triggered by continued pressure on the current account which leads to unsustainable currency depreciation; for example, the IMF-imposed austerity in the aftermath of the East-Asian financial crisis greatly protracted recovery (Stiglitz, 2002). As the national income identity proves, recovery from financial crises necessitates an increase in the government deficit to offset losses in private savings triggered by a currency crisis. Consequently, TDH-derived policies are an anathema to the needs of an economy recovering from a financial crisis.

It is important to note that this argument in no way suggests that governments should always pursue expansionary fiscal policy. It only provides a more empirically and theoretically rigorous framework to help clarify the sovereignty and fiscal space available to policymakers whilst pursuing domestic policy objectives without risking unsustainable foreign sector movements. A discussion of optimal fiscal and monetary policy would require other considerations such as the real resource constraint of the economy, which are beyond the scope of this study.

## 2A – Modelling the RER Balance: Framework & Interpretation

We draw from an income definition for the Quantity Theory of Money (QTM) (Friedman, 1970), applied to an empirically validated monetary interpretation of exchange rates (Bilson, 1978) to formulate a theoretical basis for the balance between growth and capital flows. Repurposing QTM to reconcile growth and currency dynamics is not without limitations; our study operates within these bounds, in the sense that it employs the proposed identity to facilitate scrutiny of (as opposed to defining) economically meaningful relationships, which are substantiated by more robust general models. We begin by formally specifying the inputs. For QTM, we get:

$$MV = PY \Rightarrow V = \frac{PY}{M} \Rightarrow V = \frac{Py}{M/n}$$

with $y = \frac{Y}{n}$ as real per-capita income (output), V as an arbitrary constant (formally, velocity of money), P as the relative price level, and M as money supply.

From Bilson's work on the monetary interpretation of exchange rates, we can use two models; the identity derived from money demand, referred to hereon as the monetary formulation:

$$\frac{M_a}{M_b} = S_{a/b} \cdot \left(\frac{Y_a}{Y_b}\right)^{\eta} \cdot e^{\epsilon(i_b - i_a)} \cdot e^{c_0 + \lambda t} \Rightarrow \frac{M_a}{M_b} \cdot \frac{Y_b}{Y_a} = S_{a/b} \cdot \left(\frac{Y_a}{Y_b}\right)^{\eta-1} \cdot e^{\epsilon(i_b - i_a)} \cdot e^{c_0 + \lambda t}$$

and a more traditional price-level identity with zero-period parity:

$$\frac{P_a}{P_b} = S_{a/b} \cdot e^{\psi_0} \cdot e^{\psi_1(i_b - i_a)} \Rightarrow \frac{M_a}{M_b} \cdot \frac{Y_b}{Y_a} = \frac{V_b}{V_a} \cdot S_{a/b} \cdot e^{\psi_0} \cdot e^{\psi_1(i_b - i_a)}$$

Where $S_{a/b}$ represents the spot rate for currency *a* per unit currency *b*, *i* is the policy/nominal interest rate, and $\lambda$ is a measure of the relative growth rate in money demand.

Given the EME focus of this study, a growth-oriented model is suitable for analysis, and we turn to Solow's work in the field (Solow, 1956) for its widespread applicability and robustness. With a per-capita output definition of the Solow model in case of a Cobb-Douglas production function:

$$y = z \cdot k^{\alpha} \Rightarrow \dot{y} = \alpha \cdot z \cdot k^{\alpha-1} = \frac{\alpha \cdot y}{k} \Rightarrow y = \frac{k \cdot \dot{y}}{\alpha}$$

In case of the small time-period of currency dynamics, a more direct relation to growth:

$$\dot{y} = \frac{dy}{dk} = \frac{dy}{y} \cdot \frac{y}{dk} \approx \frac{g \cdot y}{\delta k} \Rightarrow y \approx \frac{k}{\alpha} \cdot \frac{g \cdot y}{\Delta k}$$

With this theoretical grounding, we compose a QTM identity for analysis as:

$$V = \frac{P}{M/n} \cdot g_y \cdot y \cdot \frac{k}{\alpha \cdot \Delta k} = \frac{P}{M} \cdot g^y \cdot Y \cdot \frac{k}{\alpha \cdot \Delta k}$$

For two currencies, we get:

$$\frac{V_a}{V_b} = \frac{P_a}{P_b} \cdot \frac{M_b}{M_a} \cdot \frac{g_a^y}{g_b^y} \cdot \frac{Y_a}{Y_b} \cdot \frac{k_a}{k_b} \cdot \frac{\alpha_b \cdot \Delta k_b}{\alpha_a \cdot \Delta k_a}$$

Equation 1

Of the price-level identity and the monetary formulation, the former is a superior estimator of exchange rates (Bilson, 1978); however, we make an assumption regarding parity for the price-level identity that is not tested here, and it is prudent to evaluate implications in both cases. By the price-level identity, we have:

$$\frac{V_a}{V_b} = \frac{P_a}{P_b} \cdot \frac{V_a}{V_b} \cdot S_{b/a} \cdot e^{\psi_0} \cdot e^{\psi_1(i_a - i_b)} \cdot \frac{g_a^y}{g_b^y} \cdot \frac{k_a}{k_b} \cdot \frac{\alpha_b \cdot \Delta k_b}{\alpha_a \cdot \Delta k_a}$$

$$\Rightarrow 1 = RER_{b/a} \cdot e^{\psi_0} \cdot \left(\frac{e^{i_a}}{e^{i_b}}\right)^{\psi_1} \cdot \frac{g_a^y}{g_b^y} \cdot \frac{k_a}{k_b} \cdot \frac{\alpha_b \cdot \Delta k_b}{\alpha_a \cdot \Delta k_a}$$

to derive an equilibrium condition:

$$g_b^y \cdot \frac{k_b \cdot e^{\psi_1 \cdot i_b}}{\alpha_b \cdot \Delta k_b} = RER_{b/a} \cdot g_a^y \cdot \frac{k_a \cdot e^{\psi_1 \cdot i_a}}{\alpha_a \cdot \Delta k_a} \cdot e^{\psi_0}$$

Equation 2

Likewise, for the monetary formulation:

$$\frac{V_a}{V_b} = \frac{P_a}{P_b} \cdot S_{b/a} \cdot \left(\frac{Y_a}{Y_b}\right)^{\eta-1} \cdot e^{\epsilon(i_a - i_b)} \cdot e^{c_0 + \lambda t} \cdot \frac{g_a^y}{g_b^y} \cdot \frac{k_a}{k_b} \cdot \frac{\alpha_b \cdot \Delta k_b}{\alpha_a \cdot \Delta k_a}$$

$$\Rightarrow \frac{V_a}{V_b} = RER_{b/a} \cdot \left(\frac{Y_a}{Y_b}\right)^{\eta-1} \cdot e^{\epsilon(i_a - i_b)} \cdot e^{c_0 + \lambda t} \cdot \frac{g_a^y}{g_b^y} \cdot \frac{k_a}{k_b} \cdot \frac{\alpha_b \cdot \Delta k_b}{\alpha_a \cdot \Delta k_a}$$

for which equilibrium occurs at:

$$g_b^y \cdot \frac{k_b \cdot e^{\epsilon \cdot i_b}}{\alpha_b \cdot \Delta k_b} \cdot \frac{Y_b^{\eta-1}}{V_b} = RER_{b/a} \cdot g_a^y \cdot \frac{k_a \cdot e^{\epsilon \cdot i_a}}{\alpha_a \cdot \Delta k_a} \cdot \frac{Y_a^{\eta-1} \cdot e^{c_0 + \lambda t}}{V_a}$$

Equation 3

Deliberate assessment of the marginal effect on an EME's eta parameter (such as that in relation to economic maturity/development) is beyond the scope of this study, but it is likely to yield results that will be valuable to decisions concerning monetary stance and the nature of intervention by a policy authority.

Bilson (1978) computed $\eta \approx 1$ for Deutsche Mark/Pound Sterling data, which makes equilibrium via the monetary formulation functionally similar to that of the price-level identity in the context of growth:

$$g_b^y \cdot \frac{k_b \cdot e^{\epsilon \cdot i_b}}{\alpha_b \cdot \Delta k_b} \cdot \frac{1}{V_b} = RER_{b/a} \cdot g_a^y \cdot \frac{k_a \cdot e^{\epsilon \cdot i_a}}{\alpha_a \cdot \Delta k_a} \cdot \frac{e^{c_0 + \lambda t}}{V_a}$$

*Equation 4*

By both identities, we may explain an economically significant balance between economies that pivots about RER. Further, by viewing inter-economy flows occurring through a stationary medium, which may be represented by the global dollar economy, we are equipped to conduct more nuanced analysis without introducing excess complexity.

## 2B – Modelling the RER Balance: Analysis & Evaluation

For the U.S. Dollar as a stationary medium, it is necessary to define what is meant by a dollar economy; although characteristics for domestic currency *x* are sufficiently captured by national statistics (particularly in case of an EME currency), the basis for the dollar is more ambiguous:

$$g_x^y \cdot \frac{k_x \cdot e^{\psi_1 \cdot i_x}}{\alpha_x \cdot \Delta k_x} = RER_{x/\$} \cdot g_\$^y \cdot \frac{k_\$ \cdot e^{\psi_1 \cdot i_\$}}{\alpha_\$ \cdot \Delta k_\$} \cdot e^{\psi_0}$$

*Equation 5*

$$g_x^y \cdot \frac{k_x \cdot e^{\epsilon \cdot i_x}}{\alpha_x \cdot \Delta k_x} \cdot \frac{1}{V_x} = RER_{x/\$} \cdot g_\$^y \cdot \frac{k_\$ \cdot e^{\epsilon \cdot i_\$}}{\alpha_\$ \cdot \Delta k_\$} \cdot \frac{e^{c_0 + \lambda t}}{V_\$}$$

*Equation 6*

$ labels characteristics of the global dollar-denominated economy, which is distinct from the U.S. economy, giving rise to a parity condition:

$$g^y_{USD} \cdot \frac{k_{USD} \cdot e^{\psi_1 \cdot i_{USD}}}{\alpha_{USD} \cdot \Delta k_{USD}} = RER_{USD/\$} \cdot g^y_\$ \cdot \frac{k_\$ \cdot e^{\psi_1 \cdot i_\$}}{\alpha_\$ \cdot \Delta k_\$} \cdot e^{\psi_0}$$

$$g^y_{USD} \cdot \frac{k_{USD} \cdot e^{\epsilon \cdot i_{USD}}}{\alpha_{USD} \cdot \Delta k_{USD}} \cdot \frac{1}{V_{USD}} = RER_{USD/\$} \cdot g^y_\$ \cdot \frac{k_\$ \cdot e^{\epsilon \cdot i_\$}}{\alpha_\$ \cdot \Delta k_\$} \cdot \frac{e^{c_0 + \lambda t}}{V_\$}$$

where the *USD/$* spot rate must always be 1:

$$g^y_{USD} \cdot \frac{k_{USD} \cdot e^{\psi_1 \cdot i_{USD}}}{\alpha_{USD} \cdot \Delta k_{USD}} = \frac{P_\$}{P_{USD}} \cdot g^y_\$ \cdot \frac{k_\$ \cdot e^{\psi_1 \cdot i_\$}}{\alpha_\$ \cdot \Delta k_\$} \cdot e^{\psi_0}$$

$$g^y_{USD} \cdot \frac{k_{USD} \cdot e^{\epsilon \cdot i_{USD}}}{\alpha_{USD} \cdot \Delta k_{USD}} \cdot \frac{1}{V_{USD}} = \frac{P_\$}{P_{USD}} \cdot g^y_\$ \cdot \frac{k_\$ \cdot e^{\epsilon \cdot i_\$}}{\alpha_\$ \cdot \Delta k_\$} \cdot \frac{e^{c_0 + \lambda t}}{V_\$}$$

Thus, we specify a model that can be generalised across economies participating in dollar-denominated trade, as:

$$g^y_x \cdot \frac{k_x \cdot e^{\psi_1 \cdot i_x}}{\alpha_x \cdot \Delta k_x} = RER_{x/\$} \cdot \frac{P_{USD}}{P_\$} \cdot g^y_{USD} \cdot \frac{k_{USD} \cdot e^{\psi_1 \cdot i_{USD}}}{\alpha_{USD} \cdot \Delta k_{USD}}$$

*Equation 7*

$$g^y_x \cdot \frac{k_x \cdot e^{\epsilon \cdot i_x}}{\alpha_x \cdot \Delta k_x} \cdot \frac{1}{V_x} = RER_{x/\$} \cdot \frac{P_{USD}}{P_\$} \cdot g^y_{USD} \cdot \frac{k_{USD} \cdot e^{\epsilon \cdot i_{USD}}}{\alpha_{USD} \cdot \Delta k_{USD}} \cdot \frac{1}{V_{USD}}$$

*Equation 8*

The motivation for employing an equilibrium condition specified in *USD* rather than *$* terms is twofold: first, the influence of the U.S. economy (and the Federal Reserve's monetary policy) on the global economy as of writing makes a direct equilibrium with *USD* characteristics more relevant than an abstract 'global' reference frame; second, the wider applicability of the characteristics of a discrete economy, as accurate statistics for global aggregates are not very robust across use cases, and the implications of isolating dollar-denominated aggregates would alone warrant an independent study.

### 3 – Discussing Growth and its role in Currency Shocks

The simultaneous presence of a growth-adjacent transmission channel and the apparent exogeneity of fiscal policy-facilitating flows implies that the capital structure of per-capita output growth is an explanatory factor for currency dynamics. This growth-capital effect is compatible with existing literature studying the relative volatility of EME exchange rates and price levels, and is validated by the distinct monetary trajectories of developing countries with varying growth paths; it also helps illustrate a latent real vector in international financial contagion, in the form of per-capita capital development.

For the illustrative case of an EME with a target per-capita GDP growth rate that is greater than that of the U.S., a hot-growth (capital-deficit) approach would seek to achieve GDP growth greater than that which may be expected for the capital development in the current period; via the RER equilibrium, this would exert downward pressure on the existing per-capita capital stock, upward pressure on the RER, and (in case of the monetary formulation) upward pressure on the velocity of money; any change to the per-capita capital stock implies substantial structural change to the economy, which will either appear as a shock (as postulated by the TDH) or be a long-term path to sustaining the target growth.

With a relief mechanism available in the form of RER, an economy that does not actively enforce a currency peg will 'vent' the shock through some combination of rise in the exchange rate (depreciation of the domestic currency) and fall in price level, which (given conducive industrial policy) will trigger a positive feedback loop of increased competitiveness in the global market. If this mechanism is blocked by the choice of exchange rate regime, a competent monetary authority will either need to intervene and make up for the expected loss in capital stock (traditionally by defending the currency, utilising monetary reserves rather than productive capital); failure to intervene effectively will (optimistically) result in a temporary price level shock (which is only truly viable in a perfectly open economy) that gradually recedes through capital flows which may be fine-tuned through monetary policy (in addition to the velocity of money, through the monetary formulation), or more commonly, a currency crisis, as was demonstrated by the East Asian Crisis of 1997-99.

In the alternate case of more a more moderate approach to growth, an EME is not entirely immune to currency shocks; while capital-neutral growth is more manageable in terms of policy requirements and the overall monetary stance of the economy (such as by regulating domestic access to capital in order to facilitate a fixed exchange rate regime), an early- or middle-stage EME will typically rely on capital flows from abroad in the form of technology and raw material imports. This makes the short run $\Delta k$ susceptible to exogenous shocks, which are gradually amplified in the transmission to growth rate through industrial policy; as a consequence, policymakers are afforded certain degree of forewarning of the imminent shock. However, such forewarning is not always useful, as all viable short-run release mechanisms require the economy to be open for international capital flows – national reserves and monetary measures will only delay the inevitable shock.

Policy decisions in the Indian economy make this case in point: an ideal of self-sufficient growth supported by protectionist policies and central planning produced a relatively stable macroeconomic environment, with strict regulation governing capital usage and international capital flows; this policy stance sustained through the 1980s, until the limitations in $\Delta k$ made it apparent that national reserves would be insufficient to absorb the imminent capital shock, at which point sweeping changes to fiscal policy and a devaluation of the Rupee allowed the Indian economy to settle as a more open EME, as implied by the equilibrium condition for closed-economy capital-neutral growth.

The capital surplus case, where an economy grows slower than the pace enabled by its capacity for capital development, is more stable; it is a balance that an EME in the contemporary sense would be unable to maintain, as the 'emerging' aspect of these economies signifies the need to build an industrial base that can generate a surplus entirely endogenous to the domestic market. This is not to say that there will be no imports, but merely that the output in the current period covers the import bill in its entirety; i.e. the policy authority in such an economy serves to smooth fluctuations, rather than directly absorb shocks. Consistent with this reasoning, the currencies of mature/developed economies are generally more stable than their EME counterparts.

      This study outlines the mechanics of pre-empting a currency crisis, to which end we test the extent of the policymaking space available in an EME, and propose an equilibrium condition with a short-run pivot about RER; there are number of relationships the balance about RER puts forth, the evaluation of which is far beyond the intended scope of this paper. Appendix IV details two aspects that we find relevant to the context of growth in the currency dynamics of a modern EME. From the preceding discussion, we gather that capital characteristics of an EME's growth trajectory are just as important as the actual actions taken to counteract currency shocks, and play an important role in informing the policy stance of competent authorities.

**Appendix I**

Data for the period 1990-2024 of the following country / monetary authority pairs has been used as a representative set for emerging markets in the empirical study.

- (Argentina) Republic of Argentina; *Central Bank of Argentina*
- (Brazil) Federative Republic of Brazil; *Central Bank of Brazil*
- (China) People's Republic of China; *People's Bank of China*
- (India) Republic of India; *Reserve Bank of India*
- (Indonesia) Republic of Indonesia; *Bank Indonesia*
- (South Africa) Republic of South Africa; *South African Reserve Bank*

We acknowledge that this list is not exhaustive, and will not be representative of an objectively defined 'emerging' market in all contexts. This selection has been motivated by criteria adapted to the needs of this study, including subjective factors such as the accessibility of data, post-cold-war economic outlook, monetary and fiscal sovereignty, and the relative importance in capturing distinct policy paradigms.

**Appendix II**

We identify the link between macroeconomic factors and interest rates as the starting point of our analysis because the validity of the Mundell-Flemming TDH framework critically depends on the financial crowding-out argument, which suggests that increased government spending will drive up demand for loanable funds and thereby increase interest rates in the economy. This argument is based on the Loanable Funds Theory (LFT) (Robertson, 1934), which itself remains a centre of academic debate amongst economists. If LFT is proven to be empirically inconsistent, we cannot derive any direct relationship between interest rates and government spending; consequently, it is hard to argue that expansionary fiscal policy can cause capital account instability or currency crises. This result will be particularly important, because a significant portion of existing literature attributes currency crises to expansionary fiscal policy. Although such assertions are theoretically sound, an issue with generalising this outcome is that nations where expansionary fiscal policy was practiced may not necessarily be exercising monetary sovereignty (e.g. they had an exchange rate peg). Exchange rate pegs (among other factors) diminish the influence central banks have on long-term interest rates which makes testing (1) and (2) redundant.

## Appendix III

| Argentina (fiscal deficit on CAD) | | | Indonesia (fiscal deficit on CAD) | | |
|---|---|---|---|---|---|
| Lag | F-Statistic | p-Value | Lag | F-Statistic | p-Value |
| 1 | 0.4743 | 0.4967 | 1 | 0.1575 | 0.6945 |
| 2 | 1.6293 | 0.2162 | 2 | 0.6832 | 0.5142 |
| 3 | 1.0431 | 0.3931 | 3 | 1.9038 | 0.1585 |
| 4 | 1.4337 | 0.2614 | 4 | 1.7751 | 0.1756 |
| 5 | 1.1035 | 0.3968 | 5 | 1.9526 | 0.1413 |
| 6 | 1.0206 | 0.4540 | 6 | 1.9338 | 0.1501 |
| 7 | 0.9274 | 0.5252 | 7 | 1.6048 | 0.2399 |
| 8 | 0.8707 | 0.5792 | 8 | 1.2207 | 0.4028 |
| 9 | 3.1525 | 0.1405 | 9 | 0.7403 | 0.6760 |

*Table 5: Granger causality (fiscal deficit on CAD) statistics for Argentina and Indonesia*

| Brazil (fiscal deficit on CAD) | | | South Africa (fiscal deficit on CAD) | | |
|---|---|---|---|---|---|
| Lag | F-Statistic | p-Value | Lag | F-Statistic | p-Value |
| 1 | 0.0534 | 0.8196 | 1 | 2.8191 | 0.1080 |
| 2 | 0.4970 | 0.6169 | 2 | 4.5451 | 0.0252 |
| 3 | 0.1924 | 0.8999 | 3 | 2.8384 | 0.0734 |
| 4 | 0.1940 | 0.9364 | 4 | 1.3821 | 0.2978 |
| 5 | 0.7804 | 0.5907 | 5 | 3.1168 | 0.0661 |
| 6 | 1.1479 | 0.4494 | 6 | 2.0856 | 0.1964 |
| 7 | 1.4230 | 0.4729 | 7 | 7.9474 | 0.0582 |

*Table 6: Granger causality (fiscal deficit on CAD) statistics for Brazil and South Africa*

| China (fiscal deficit on CAD) | | |
|---|---|---|
| Lag | F-Statistic | p-Value |
| 1 | 0.0343 | 0.8546 |
| 2 | 0.2494 | 0.7817 |
| 3 | 0.5956 | 0.6264 |
| 4 | 1.2883 | 0.3214 |
| 5 | 1.6836 | 0.2189 |

*Table 7: Granger causality (fiscal deficit on CAD) statistics for China*

**Appendix IV**

**Relative Capital Neutrality**: we introduce the idea of a shared stationary medium, but do not evaluate its full implications in this study; every economy will have a unique capital-neutral growth rate, but trade between economies requires this neutral rate to be constant between economies; inter-economy capital flow transactions may be reconciled by reviewing the balance about RER from the perspective of this stationary medium, where the neutral rate for all economies is an absolute reference point; a weak analogy may be made to the parity between two frames of reference in motion relative to a stationary observer, with observations reconciled by means of the Lorentz transformation;

**Debt-financed Capital**: domestic debt financing conserves capital within an economy, but international debt financing will provide a one-time boost to $k$ at the cost of $\Delta k$ in subsequent periods; this has distinct implications for financing both private enterprise, and government activity; private enterprise raises finance from abroad to fund capital investment that is expected to provide direct economic output to facilitate the payment of debt with normal profit, implying that any $\Delta k$ is offset by an effect on growth rate; international financing for government policy (macroeconomically) expects the aggregate outcome of the investment to pay for the debt, but debt payments are made irrespective of return; government financing pledges $\Delta k$ to deliver a compensating growth rate, which has far-reaching ramifications for the choice of government participation in a capital-deficit and capital neutral EME;

### Index of Figures & Tables

**Data Sources & Work Cited**